\PassOptionsToPackage{unicode}{hyperref}
\PassOptionsToPackage{hyphens}{url}
\documentclass[
]{article}
\usepackage{xcolor}
\usepackage{amsmath,amssymb}
\usepackage{iftex}
\ifPDFTeX
  \usepackage[T1]{fontenc}
  \usepackage[utf8]{inputenc}
  \usepackage{textcomp} 
\else 
  \usepackage{unicode-math} 
  \defaultfontfeatures{Scale=MatchLowercase}
  \defaultfontfeatures[\rmfamily]{Ligatures=TeX,Scale=1}
\fi
\usepackage{lmodern}
\ifPDFTeX\else
\fi
\IfFileExists{upquote.sty}{\usepackage{upquote}}{}
\IfFileExists{microtype.sty}{
  \usepackage[]{microtype}
  \UseMicrotypeSet[protrusion]{basicmath} 
}{}
\makeatletter
\@ifundefined{KOMAClassName}{
  \IfFileExists{parskip.sty}{%
    \usepackage{parskip}
  }{
    \setlength{\parindent}{0pt}
    \setlength{\parskip}{6pt plus 2pt minus 1pt}}
}{
  \KOMAoptions{parskip=half}}
\makeatother
\usepackage{longtable,booktabs,array}
\usepackage{calc} 
\usepackage{etoolbox}
\makeatletter
\patchcmd\longtable{\par}{\if@noskipsec\mbox{}\fi\par}{}{}
\makeatother
\IfFileExists{footnotehyper.sty}{\usepackage{footnotehyper}}{\usepackage{footnote}}
\makesavenoteenv{longtable}
\usepackage{graphicx}
\makeatletter
\newsavebox\pandoc@box
\newcommand*\pandocbounded[1]{
  \sbox\pandoc@box{#1}%
  \Gscale@div\@tempa{\textheight}{\dimexpr\ht\pandoc@box+\dp\pandoc@box\relax}%
  \Gscale@div\@tempb{\linewidth}{\wd\pandoc@box}%
  \ifdim\@tempb\p@<\@tempa\p@\let\@tempa\@tempb\fi
  \ifdim\@tempa\p@<\p@\scalebox{\@tempa}{\usebox\pandoc@box}%
  \else\usebox{\pandoc@box}%
  \fi%
}
\def\fps@figure{htbp}
\makeatother
\usepackage{bookmark}
\IfFileExists{xurl.sty}{\usepackage{xurl}}{} 
\hypersetup{
  hidelinks,
  pdfcreator={LaTeX via pandoc}}

\author{}
\date{}

\begin{document}

Corresponding author: Shengyu Gao (Email:
\url{ShengyuGao0327@outlook.com})

\textbf{Making Smell Actionable in Design: Lexicon, Movement Patterns, and
Gesture Cards}

Anran Feng\textsuperscript{1}, Shengyu Gao\textsuperscript{2}*, Xingyu
Zhao\textsuperscript{2}, Zhenyu Qian\textsuperscript{1}, Jingrui
An\textsuperscript{3}

1.Patti \& Rusty Rueff School of Design, Art, and Performance, Purdue
University

2.Birmingham Institute of Fashion and Creativity, Birmingham City
University

3.School of Digital Technology and Innovation Design, Jiangnan
University

Smell is powerful in design, yet it often remains vague and difficult to
work with in studios and classrooms. While olfactory HCI has produced
devices, toolkits, and scenario-based applications, designers still lack
teachable, shareable resources that link smell to bodily action. This
paper presents a three-phase research program that makes smell
actionable by moving from lived smell experiences to movement patterns
and, finally, to gesture resources. We study olfactory--motor mappings
and package the results as teachable design representations through a
progressively narrowing method: (1) extracting smell--verb--intent
relations from user narratives; (2) capturing and clustering embodied
enactments into standardized upper-limb trajectory families; and (3)
translating these trajectories into gesture card representations for
design teaching and ideation through design workshops. We present a
three-phase translational framework and empirically derived
smell--gesture correspondences that structure the
smell--intent--trajectory--gesture pathway, together with reusable
assets (lexicon, trajectory atlas, gesture tables, and cards) that
support teaching, reuse, and adaptation across contexts.

\textbf{Keywords} -- Design representations, Gesture cards, Movement
patterns, Olfactory interaction, Smell lexicon

\textbf{Relevance to Design Practice} -- This work offers designers an
exploratory framework for translating smell experiences into embodied
representations that can be shared, compared, and adapted in studio
settings. The resulting lexicon, trajectory families, and gesture cards
are designed to support ideation, teaching, and reflective critique,
while remaining open to reinterpretation across projects and contexts.

\section{1 INTRODUCTION}\label{introduction}

Design practice needs concrete ways to make multi-sensory and embodied
experience shareable and actionable in studios and classrooms. Embodied
and affect-based theories argue that cognition is grounded in dynamic
body--world coupling, where interaction, context, and felt qualities
shape judgment and action (Barsalou 2008; Greifeneder et al. 2011;
Schwarz and Clore 1983; Slovic et al. 2007; Wilson 2002; Dourish 2001;
Hook 2018). In this context, olfactory HCI has advanced devices and
design spaces (e.g., wearables, proximal delivery, and mapping
frameworks) and called for more systematic design and evaluation of
smell as an interaction modality (Dobbelstein et al. 2017; Maggioni et
al. 2020; Wang et al. 2020; Amores et al. 2017; Brooks et al. 2023;
Obrist et al. 2016), yet studio-ready resources that connect smell to
enactable bodily forms remain underdeveloped, limiting what teams can
teach, critique, and reuse in practice (Wang et al. 2024). We address
this gap by building a smell--intent lexicon from user stories,
capturing and analyzing upper-limb enactments to derive trajectory
patterns and form hints, and synthesizing reusable Gesture Cards plus a
lightweight workshop protocol to support teaching, sharing, and
adaptation across projects and studios.This study is exploratory in
nature, the resulting representations are intended as design scaffolds
in the sense proposed by Höök and Löwgren (2012), supporting ideation
and teaching rather than constituting validated tools ready for
deployment.

\section{2 RELATED WORK}\label{related-work}

\subsection{2.1 Embodied Cognition and
Olfaction}\label{embodied-cognition-and-olfaction}

Embodied cognition frames thinking as grounded in the dynamic coupling
of body and environment, where sensation, movement, and context shape
mental activity (Barsalou 2008; Dourish 2001; Wilson 2002). Within this
view, olfaction offers a direct entry point to perception--action
relations: the PROUST hypothesis links smell to spatial--temporal
reconstruction and behavioral adjustment (Jacobs 2023), and
Brain-imaging evidence suggests location-sensitive representations that
support spatial processing of smell (Poo et al. 2022). Smell also
carries affective and metaphorical value for judgment and decision
making (Herz 2016) and for categorization and social evaluation at the
intersection of language, culture, and biology (Majid 2021). Crossmodal
work further reports relatively stable correspondences between smell and
color, touch, and shape (Spence 2020; Dematte et al. 2006; Hanson-Vaux
et al. 2013), supporting the idea that people can learn smell-based
regularities and reuse them in new situations. Finally, olfactory input
can bypass thalami relay and engage limbic circuits, suggesting fast
pathways that may shape action and choice before conscious awareness
(Wilson 2002).

\subsection{2.2 Smell in Design Practice: Applications and
Implications}\label{smell-in-design-practice-applications-and-implications}

Smell matters in design because it engages memory and feeling and aligns
with embodied interaction (Herz 2016; Dourish 2001). Evidence on
crossmodal links and affective valence suggests that smell can support
learnable design regularities and influence evaluation and choice
(Spence 2020; Dematte et al. 2006; Hanson-Vaux et al. 2013; Puska et al.
2018), potentially shaping orientation and action through fast pathways
(Wilson 2002).

To lower barriers in practice, prior work has proposed use cases and
toolchains: early arguments for olfactory output in HCI and systematic
use beyond novelty (Kaye 2004) have been extended by programmable and
teachable toolkits and prototyping approaches (Maggioni et al. 2019; Lei
et al. 2022; Brooks and Lopes 2023; Li et al. 2025), alongside reviews
calling for more systematic design and evaluation (Obrist et al. 2016)
and design-space work mapping opportunities and dimensions across
chemical, emotional, spatial, and temporal factors (Obrist et al. 2014;
Maggioni et al. 2020).

Applications demonstrate how olfactory resources play out across
contexts, including wearable and near-face systems, immersive
interactions, collaboration, safety alerts, and VR experience shaping
(Dobbelstein et al. 2017; Wang et al. 2020; Niedenthal et al. 2023;
Mehrotra et al. 2022; Dmitrenko et al. 2017; Flavián et al. 2021).
Evaluation practices are also broadening, from experience-centered
metrics and frameworks to standardized olfactory tests, implicit
association methods, and posture/balance effects relevant to embodied
scenarios (Lin et al. 2024; Hummel et al. 1997; Greenwald et al. 1998;
Hanson-Vaux et al. 2013; Li et al. 2008; Oleszkiewicz et al. 2023;
Schreuder et al. 2014).

\textbf{Gap and positioning.}Prior work has advanced olfactory
techniques and design spaces, but it rarely links smell to teachable
bodily action for studio use. We respond with portable resources---a
smell--intent lexicon, recurrent movement patterns, and Gesture Cards
with a lightweight workshop protocol---to make smell more actionable in
practice (Höök and Löwgren 2012; Dourish 2001).

\section{3 Overview of Studies}\label{overview-of-studies}

We carried out three connected phases to build studio-ready design
representations for smell and bodily expression: Phases 1 and 2 produce
empirically grounded resources (lexicon and trajectory families); Phase
3 translates these into gesture card representations for teaching and
ideation.(see in Fig. 1).

\begin{center}
\includegraphics[width=\linewidth,height=0.9\textheight,keepaspectratio,alt={截屏2026-01-06 下午3.31.16}]{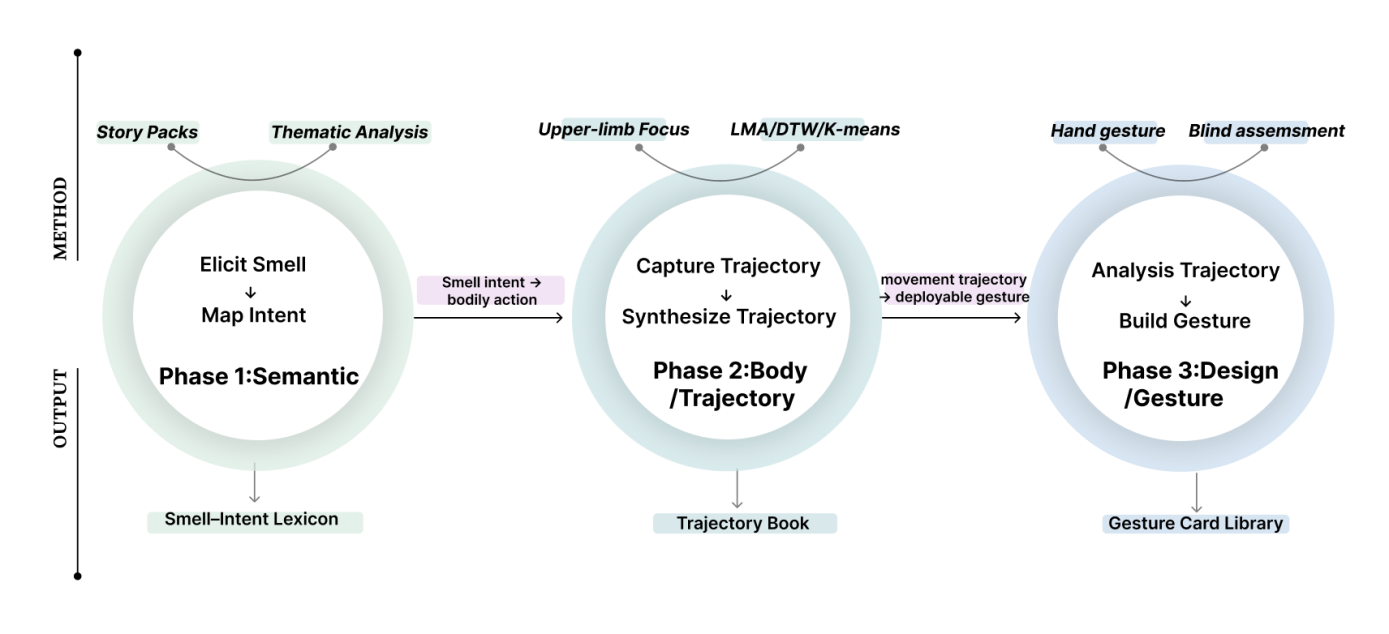}
\end{center}
Fig.1. Study road map

\emph{\textbf{Phase 1: From smell to verb and intent}}

We began with participants' words and short stories for five smells
(grape, citrus, mint, lavender, coffee). For each smell, we extracted
the Top-8 action verbs through frequency counts with synonym merging,
then checked face validity with three designers. We consolidated verbs
into seven operational intents (Release, Sustain, Amplify, Attenuate,
Blend/Switch, Locate, Transfer), forming the semantic base for later
phases (Maggioni et al. 2020; Obrist et al. 2014). Texts were organized
with thematic analysis and affinity diagramming (Braun and Clarke 2006;
Holtzblatt and Beyer 1997).

\emph{\textbf{Phase 2: From intent to trajectory}}

Lay participants enacted and verbalized selected verbs; movements were
recorded through multi-camera video and skeleton tracking using
representative Kinect/OpenPose pipelines (Coppola and Slotnick 2018; Cao
et al. 2017; Shotton et al. 2011). We also invited lead users to
complete the same procedure (Hippel 1986). From segmented sub-movements,
we extracted features and applied k-means (k by silhouette) to test
consistency across trained movers and lay participants and derive
high-consistency intent--trajectory links (Rousseeuw 1987). We aligned
semantics and geometry using Laban Movement Analysis and checked
``same-form'' criteria through Procrustes alignment and dynamic time
warping (Bartenieff and Lewis 2013; Dryden and Mardia 2016; Guest 2013;
Müller 2007). Outputs were a Trajectory Book with exemplars and a
smell--intent--trajectory table for Phase 3.

\emph{\textbf{Phase 3: From trajectory to gesture}}

Designers viewed and reproduced trajectory exemplars, made minimal
adjustments, and finalized 2D renderings for communication and teaching.
Each result was packaged as a Gesture Card. We then ran a formative
check (blind ratings and redo) to assess clarity and stability. The
outcome is a set of cards and a lightweight workshop protocol for
studios and classrooms.

\section{3 PHASE 1 --- From smell to verbs and
intents}\label{phase-1-from-smell-to-verbs-and-intents}

\subsection{3.1 Aim}\label{aim}

Build a smell--verb/intent base from participants' stories as semantic
input for later phases.

\subsection{3.2 Participants \& materials}\label{participants-materials}

N = 28. Five commercial samples (grape, citrus, mint, lavender, coffee)
were presented at 0.5\% concentration, following design dimensions in
Opportunities for Odor and SMELL SPACE (Maggioni et al. 2020; Obrist et
al. 2014). Sessions took place in a quiet, smell-neutral room;
participants wore a blindfold. Smells were randomized with ≥30 s
inter-trial intervals.

\subsection{3.3 Procedure}\label{procedure}

For each smell, participants sniffed for \textasciitilde5 s, recalled or
imagined a personal scene, and narrated one concrete story. We recorded
audio/video and transcribed the recordings (SRT/text), removing filler
words and adding minimal connectives only to clarify action sequences.
Using Python with Jieba, we segmented the text and applied
part-of-speech filtering to extract candidate action verbs, then merged
synonyms to produce a Top-8 verb list per smell. Two assistants removed
non-actions, consolidated close variants, and mapped high-frequency
verbs to seven operational intents (Release, Sustain, Amplify,
Attenuate, Blend/Switch, Locate, Transfer).

\subsection{3.3 Key Finding}\label{key-finding}

\textbf{From user stories to building the scene and action lexicon.}
Users' descriptions of smells are not isolated sensory fragments. They
often appear as scene associations and narrative episodes. For example,
when describing grape, users recall events such as ``press the shower
gel, work up foam, and pat it onto the body,'' ``wandering in a
streetwear shop to buy clothes,'' or ``peel the chilled grapes before
eating.'' These fragments reflect a tight connection between smell
memories and event memories. Based on this insight, we conducted open
coding on 140 real-life smell-experience narratives, extracted actions
and scenarios, and built a high-frequency action-verb lexicon with
scenario/location categories (See Table 1)

Table 1. Smell category ---typical scenarios ---locations
---high-frequency action-verb lexicon

\begin{center}
\includegraphics[width=\linewidth,height=0.9\textheight,keepaspectratio,alt={截屏2026-01-25 下午3.24.20}]{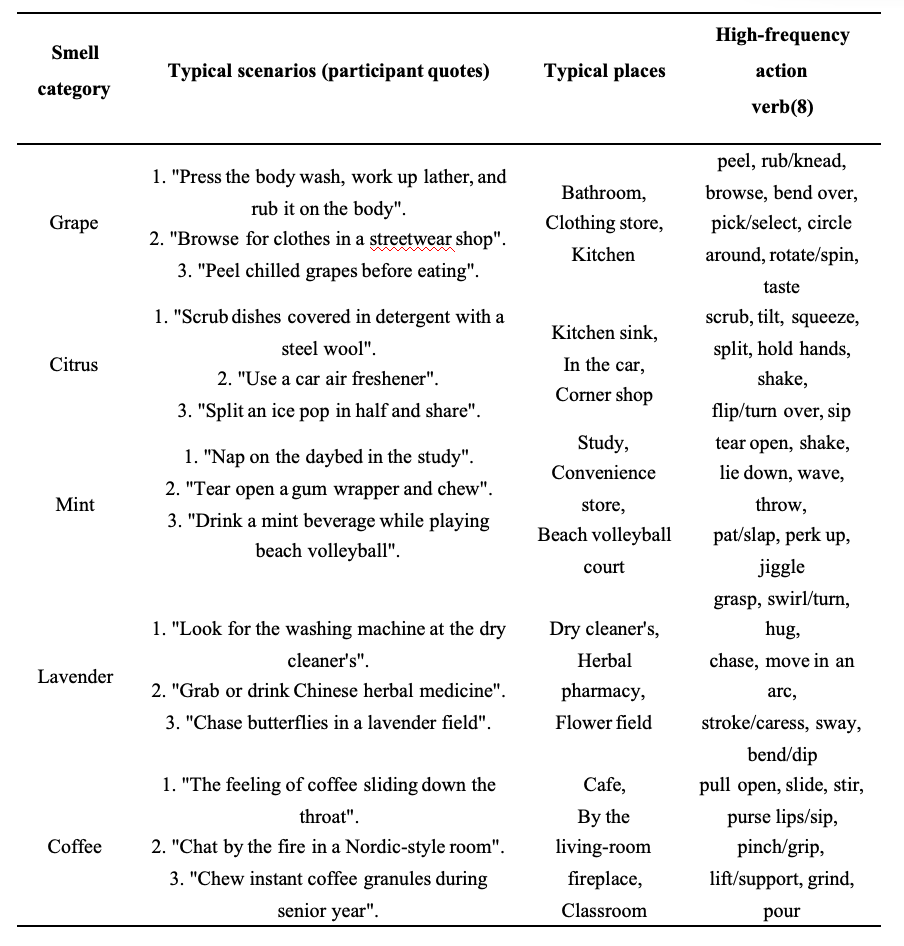}
\end{center}

On this basis, we treat ``what users want the smell to do'' as a key
factor shaping action choice. Table 1 shows clear tendencies linking
smell categories to recurrent actions (e.g., citrus with
scrubbing/squeezing; mint with more relaxed postures). Together, these
co-occurring story fragments and actions form olfactory cognitive
schemas that connect to concrete behaviors and motivate action-linked
intents, consistent with embodied accounts of smell in which action
associations arise alongside recall (Barsalou 2008).

\textbf{From the Action-Verb Lexicon to reclassifying smell operation
intents.} We next examined the intent behind actions by asking what
change users expect the smell to undergo when performing an action. We
convened a focus group of five interaction designers to conduct a
secondary analysis of the high-frequency verbs. The designers consulted
existing action-classification ideas in olfactory interaction (Li et al.
2025) but did not adopt the ``operability/mobility'' framework (Li et
al. 2025). Instead, through iterative discussion grounded in our corpus,
they focused on users' subjective goals and proposed a classification
based on ``what change users expect the smell to have.'' We refer to
these as smell operation intents, which guide the choice of specific
actions. This intent-first stance aligns with user-defined gesture
practice where intents/commands structure elicitation and analysis
(Vatavu and Wobbrock 2015; Vatavu and Wobbrock 2022; Wobbrock et al.
2009).

Table 2. Smell Operation Intents: Category-Intent-Representative Action
Mapping

\begin{center}
\includegraphics[width=\linewidth,height=0.9\textheight,keepaspectratio,alt={截屏2026-01-25 下午3.28.25}]{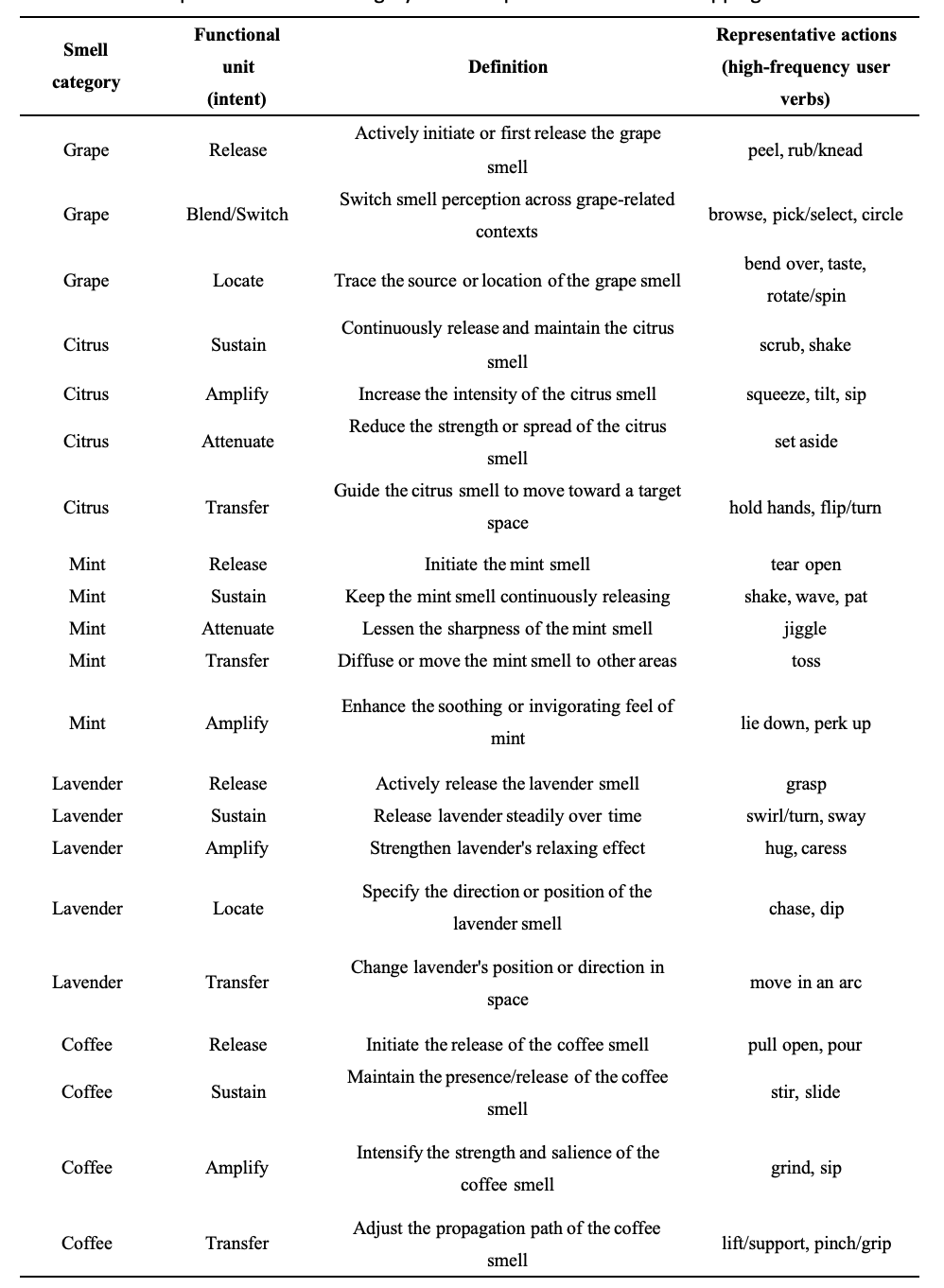}
\end{center}

We defined the intent categories in three steps and arrived at seven
non-overlapping functional units. First, designers reviewed the 38
high-frequency verbs and the motivations implied in user narratives.
Second, each designer independently proposed intent categories and the
group then converged through discussion. Third, after multiple rounds,
we finalized seven functional units with explicit definitions: Release,
Sustain, Amplify, Attenuate, Blend/Switch, Locate, and Transfer. This
produced an integrated categorization linking smell category,
action-level functional intents, and representative actions (Table 2).
These seven intents serve as the Smell-Verb-Intent output of Phase 1 and
prepare the ground for movement capture and evaluation in later phases
(Hook 2018; Vatavu and Wobbrock 2015; Vatavu and Wobbrock 2022; Wobbrock
et al. 2009).

\section{4 PHASE 2: INTENT→TRAJECTORY}\label{phase-2-intenttrajectory}

\subsection{4.1 Aim}\label{aim-1}

This phase captures upper-limb enactments of the Phase 1 intents,
translates them into parameterized trajectories through geometric and
semantic alignment (LMA, Procrustes/DTW), and clusters them into
trajectory families for Phase 3.

4.2 Participants \& Equipment

\textbf{Lay Participants:} 15 individuals (age 18--26; balanced gender)
with no formal movement training.

\textbf{Trained Movers:} 8 dancers with at least 5 years of formal
movement training. Their specialization is modern dance (Hippel 1986)
Recording

\textbf{Equipment:} For the frontal view, the setup comprised a Kinect
v2 depth sensor and a Sony A7C II; for the side view, a DJI Pocket 3.
Data were captured with synchronized dual-camera recording (1080p/60
fps) and Kinect v2 skeleton tracking at 30 fps (25 joints), following
(Cao et al. 2017; Coppola and Slotnick 2018; Oleszkiewicz et al. 2023;
Shotton et al. 2011; Shotton et al. 2011). For each smell category,
top-8 verbs from Phase 1 were presented on a printed card.

\subsection{4.3 Procedure}\label{procedure-1}

\textbf{Step 1 --- Verb selection.} All participants were blindfolded
and smelled each smell for 5 s (randomized), then selected 3--5 verbs
from the top-8 verb cards that best matched their perceived sensation.

\textbf{Step 2 --- Movement enactment (lay participants).} For each
selected verb, participants performed a spontaneous upper-body gesture
(8--10 s per verb). Movements were recorded using two cameras and
Kinect(see Fig. 2).

\textbf{Step 3 --- Improvisation and amplification (trained movers).}
Trained movers produced a 60s improvisation per smell while verbalizing
creative intentions. Researchers segmented each improvisation into 3--4
sub-actions, which movers then re-enacted systematically with varied
intensity (Sebanz and Shiffrar 2009)(see Fig. 2).

\textbf{Step 4 --- Annotation and analytic focus.} During collection,
experts performed full-body enactments to check movement-quality
boundaries. For analysis, we focused on upper-limb joints (shoulder,
elbow, wrist) and end-effector paths. Trunk/lower limbs mainly support
anticipatory postural adjustments and can be separated from end-effector
geometry (Aruin and Latash 1995; Wobbrock et al. 2009). Lower-limb/gait
tracking is also more sensitive to occlusion and scene factors in
Kinect/OpenPose than upper-limb reaching (Clark et al. 2019; Clark et
al. 2012; Guess et al. 2022; Springer and Seligmann 2016).

\textbf{Step 5 --- Features, alignment, and clustering.} We computed
duration, path length, peak acceleration, curvature complexity, and
energy distribution for upper-limb trajectories; clustered them with
k-means (k by silhouette index) (Rousseeuw 1987); aligned trajectories
with Procrustes and DTW; and used LMA (Body/Effort/Shape/Space) labels
as semantic anchors (Bartenieff and Lewis 2013; Dryden and Mardia 2016;
Guest 2013; Müller 2007).

\begin{center}
\includegraphics[width=0.49\linewidth,height=0.9\textheight,keepaspectratio]{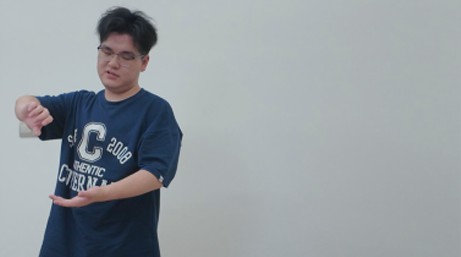}%
\hfill
\includegraphics[width=0.49\linewidth,height=0.9\textheight,keepaspectratio]{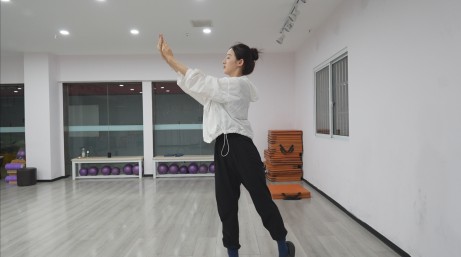}
\end{center}

Fig. 2. Example of the experimental setup: (a) lay participants; (b)
trained movers (modern dancer).

\textbf{Step 6 --- Reliability and output.} Two trained coders applied
LMA labels to each sub-action; agreement was Cohen's κ \textgreater{}
0.80 (Cohen 1960; McHugh 2012). Outputs were trajectory families
T01--T13 with corresponding semantic labels for Phase 3 design
synthesis.

\subsection{4.4 Key Finding}\label{key-finding-1}

We report Phase-2 at three levels: (1) raw data determination \&
preprocessing, (2) trajectory consistency analysis, and (3) linking
smell--intent--action--trajectory (SVIT).

\textbf{Raw data determination and pre-processing.} Lay participants and
trained movers selected largely the same actions for each smell--intent
pair; only a few items and two extra actions differed, which we
reconciled into a single action set for all subsequent analyses.

Upper-body movement was recorded with Kinect v2 (30 fps, 25 joints) and
stored as CSV files with columns Time, JointName, X, Y, Z. For all
trajectories, cleaning operated per joint and per coordinate and
comprised three steps:

\emph{Gap imputation (linear interpolation).} For short missing segments
between valid timestamps\(t_{l}\)\textless{}\(t\)\textless{} \(t_{r}\),
with joint positions p(\(t_{l}\)), p(\(t_{r}\)) ∈ R3 (CameraSpace:X,Y,Z
in meters), we impute:

{\def\LTcaptype{none} 
\begin{longtable}[]{@{}
  >{\raggedright\arraybackslash}p{(\linewidth - 4\tabcolsep) * \real{0.2450}}
  >{\raggedright\arraybackslash}p{(\linewidth - 4\tabcolsep) * \real{0.4940}}
  >{\raggedright\arraybackslash}p{(\linewidth - 4\tabcolsep) * \real{0.2610}}@{}}
\toprule\noalign{}
\endhead
\bottomrule\noalign{}
\endlastfoot
&
\(p(t) = p(t_{l}) + \frac{t - t_{l}}{t_{r -}t_{l}}(p(t_{r}) - p(t_{l}))\)
& (1) \\
\end{longtable}
}

applied component-wise to X,Y,Z. We only interpolate short gaps caused
by brief occlusions/jitter; longer gaps are excluded from analysis.

\emph{Robust temporal smoothing through median filtering (kernel = 5).}
To suppress impulsive spikes and single-frame jumps, we apply a 5-point
median filter independently to each coordinate stream:

{\def\LTcaptype{none} 
\begin{longtable}[]{@{}
  >{\raggedright\arraybackslash}p{(\linewidth - 4\tabcolsep) * \real{0.2450}}
  >{\raggedright\arraybackslash}p{(\linewidth - 4\tabcolsep) * \real{0.4940}}
  >{\raggedright\arraybackslash}p{(\linewidth - 4\tabcolsep) * \real{0.2610}}@{}}
\toprule\noalign{}
\endhead
\bottomrule\noalign{}
\endlastfoot
&
\({x\tilde{}}_{t} = median\{ x_{t - 2},x_{t - 1},x_{t},x_{t + 1},x_{t + 2}\}\)
& (2) \\
\end{longtable}
}

\emph{Normalization.} Finally, we resampled trajectories to a common
length L, centred them (removing translation), and scaled amplitudes by
the 95th-percentile path radius to support later similarity analysis.

\emph{Output.} The cleaned data keep the canonical columns Time,
JointName, X, Y, Z and feed the intent-trajectory analysis database.

\textbf{(2)Trajectory Consistency Analysis.} Consistency analysis. Using
the analysis database built in Step 1, we applied Dynamic Time Warping
(DTW) with a 2D Euclidean point-wise cost and a 15\% Sakoe--Chiba
bandwidth to prevent over-alignment. For each action, the system
computes DTW distances between the trained movers exemple and all lay
participants samples, then converts distances into similarity scores
through an exponential transform (following Barsalou, 2008).

We export (i) the full distance/similarity matrix, (ii) a thresholded
set with s ≥ 0.60, and (iii) per-action top-K matches to support
subsequent consistency analyses and summary statistics. This procedure
accommodates tempo differences while yielding an intuitive scalar
measure of geometric agreement between trained moversand participant
trajectories (DTW-based trajectory similarity; bandwidth 15\%; threshold
s ≥ 0.60 (Barsalou 2008)).

\emph{Intra-trained-mover consistency (prototype and T labels).} Within
each smell--intent--action cell, we compute all-pairs DTW among the
eight trained movers samples and report the within-group median
similarity \(\widetilde{s}\) and \(\widetilde{s}\). The median
trajectory is taken as the trained-mover prototype for that action and
assigned a trajectory-family label T01--T13 (for reuse and retrieval).
This step does not change any similarity statistics.

\emph{Trained-mover--lay-participant consistency (teachability).} Using
the trained-mover prototype as the reference, we compute DTW-based
similarity to all lay-participant samples, yielding for each action the
median similarity \(\widetilde{s}\) and the best-case similarity
\(s_{\max}\). As shown in Table 3, lay participants and trained movers
achieve moderate-to-high similarity on the same actions: global median
\(\widetilde{s}\) = 0.745, mean \(\bar{s}\) = 0.719,
P\textsubscript{25}--P\textsubscript{75} = 0.616--0.832 (IQR = 0.216),
range = 0.235-0.954. Top matches reach s = 0.86-0.89. Residual
discrepancies mainly reflect differences in amplitude and tempo. Coder
agreement on LMA labels is high (Cohen's κ \textgreater{} 0.80; see
Procedure). Table 4 lists higher-consistency smell--action pairs, ranked
by median DTW-based similarity between trained-mover prototypes and
lay-participant executions. Table 5 lists lower-consistency
smell--action pairs, ranked by the same metric; representative
lower-consistency actions include bend over, stroke/caress, tilt, and
circle around.

Table 3. Summary Statistics of NN Pairs

\begin{center}
\includegraphics[width=\linewidth,height=0.9\textheight,keepaspectratio,alt={截屏2026-01-25 下午3.28.56}]{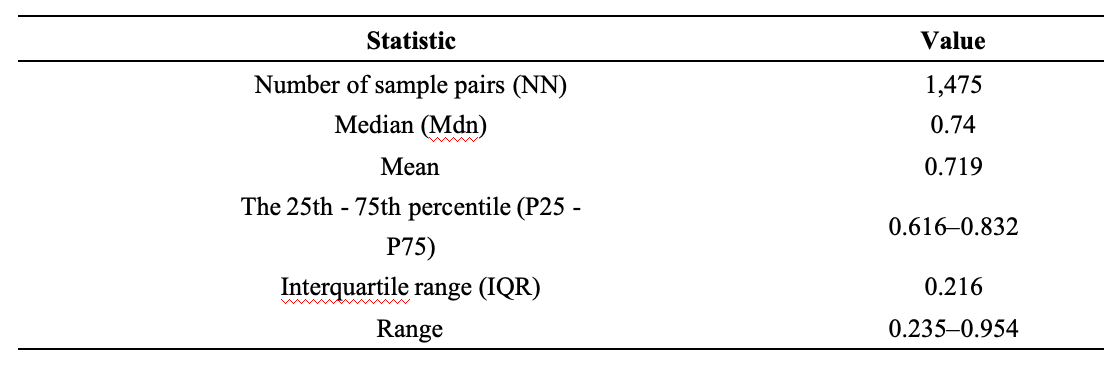}
\end{center}

Table 4. Higher-consistency smell--action pairs ranked

\begin{center}
\includegraphics[width=\linewidth,height=0.9\textheight,keepaspectratio,alt={截屏2026-01-25 下午3.29.07}]{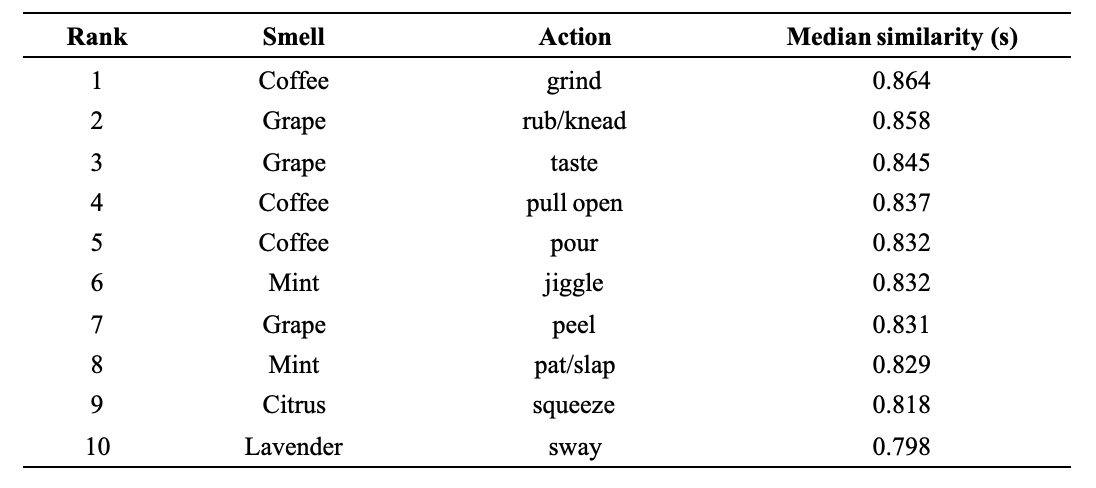}
\end{center}

Table 5. Lower-consistency smell--action pairs ranked

\begin{center}
\includegraphics[width=\linewidth,height=0.9\textheight,keepaspectratio,alt={截屏2026-01-25 下午3.29.25}]{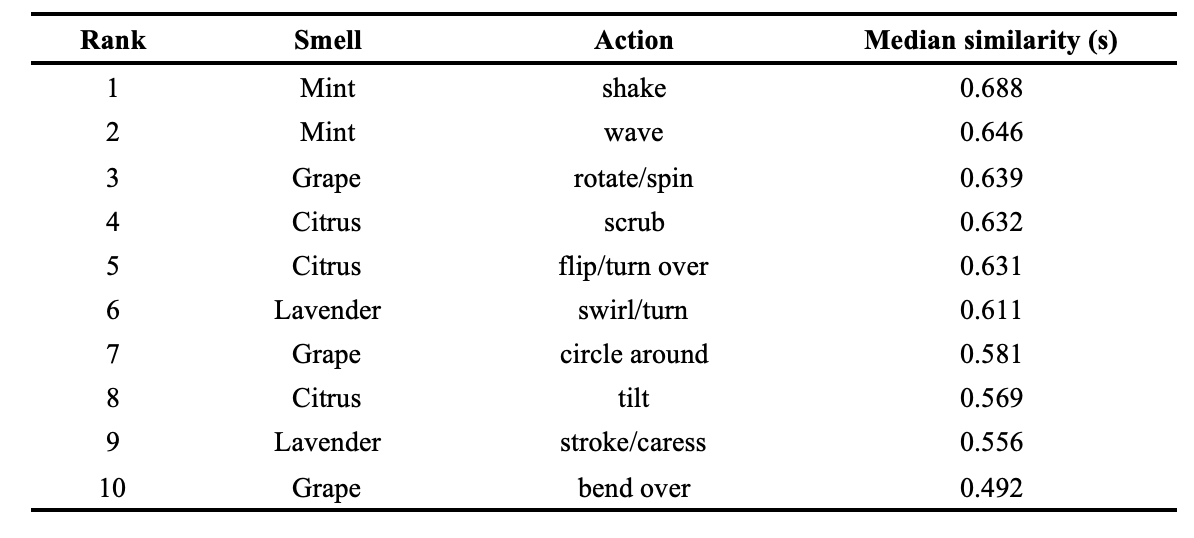}
\end{center}

\begin{enumerate}
\def\labelenumi{(\arabic{enumi})}
\setcounter{enumi}{2}
\item
  \textbf{Generating the Smell--Intent--Verb--Trajectory (SVIT) graph.}
  To construct the SVIT data graph, we aggregate and filter the smell,
  intent, action entries from Step 2 that showed higher consistency
  (Table 4). Using the trained-mover prototype as the reference, we
  proceed as follows:
\end{enumerate}

For each smell \(o\) and intent \(i\), let \(A(o,i)\) be the set of
actions collected in Step 2. For each action \(A(o,i)\), define the
DTW-based similarity multiset

{\def\LTcaptype{none} 
\begin{longtable}[]{@{}
  >{\raggedright\arraybackslash}p{(\linewidth - 4\tabcolsep) * \real{0.2450}}
  >{\raggedright\arraybackslash}p{(\linewidth - 4\tabcolsep) * \real{0.4940}}
  >{\raggedright\arraybackslash}p{(\linewidth - 4\tabcolsep) * \real{0.2610}}@{}}
\toprule\noalign{}
\endhead
\bottomrule\noalign{}
\endlastfoot
& \(S_{a} = \left\{ {S_{a}}^{(j)} \right\}\) & (3) \\
\end{longtable}
}

where each \({S_{a}}^{(j)}\) is the similarity between the trained
moversprototype and a lay participantssample under the same \(o\),
\(i\), \(a\) (DTW with 2D Euclidean cost and bandwidth constraint;
similarity as define in Step 2). Define the intent score as

{\def\LTcaptype{none} 
\begin{longtable}[]{@{}
  >{\raggedright\arraybackslash}p{(\linewidth - 4\tabcolsep) * \real{0.2450}}
  >{\raggedright\arraybackslash}p{(\linewidth - 4\tabcolsep) * \real{0.4940}}
  >{\raggedright\arraybackslash}p{(\linewidth - 4\tabcolsep) * \real{0.2610}}@{}}
\toprule\noalign{}
\endhead
\bottomrule\noalign{}
\endlastfoot
& \(IntentScore(o,i) = \max_{a \in A(o,i)}{median\left( S_{a} \right)}\)
& (4) \\
\end{longtable}
}

For each smell \(o\), rank intents \(i\) by \(IntentScore(o,i)\) and
keep the Top-3 intents. For each retained intent \(i\), choose the
representative action.

{\def\LTcaptype{none} 
\begin{longtable}[]{@{}
  >{\raggedright\arraybackslash}p{(\linewidth - 4\tabcolsep) * \real{0.2797}}
  >{\raggedright\arraybackslash}p{(\linewidth - 4\tabcolsep) * \real{0.4216}}
  >{\raggedright\arraybackslash}p{(\linewidth - 4\tabcolsep) * \real{0.2987}}@{}}
\toprule\noalign{}
\endhead
\bottomrule\noalign{}
\endlastfoot
&
\(a^{\bigstar} = \left\{ \arg\max_{a \in A(o,i)}{median\left( S_{a} \right)}\  \right\}\)
& (5) \\
\end{longtable}
}

Ties are broken by \(mean(S_{a})\) or by lexicon consistency. For each
representative action \(a^{\bigstar}\), record: the top-match pair
achieving \(s_{\max} \in S_{a^{\bigstar}}\) (for
visualization/traceability), and the corresponding trajectory-family
label \(T_{x}\) (from Step 2) for reuse and retrieval.

For every representative item, we generated an expert 2.5D trajectory
rendering and a 2D plan-view path from the original data (see Fig. 3;
further details in Methods).

\begin{center}
\includegraphics[width=\linewidth,height=0.9\textheight,keepaspectratio,alt={截屏2025-11-24 下午4.06.09}]{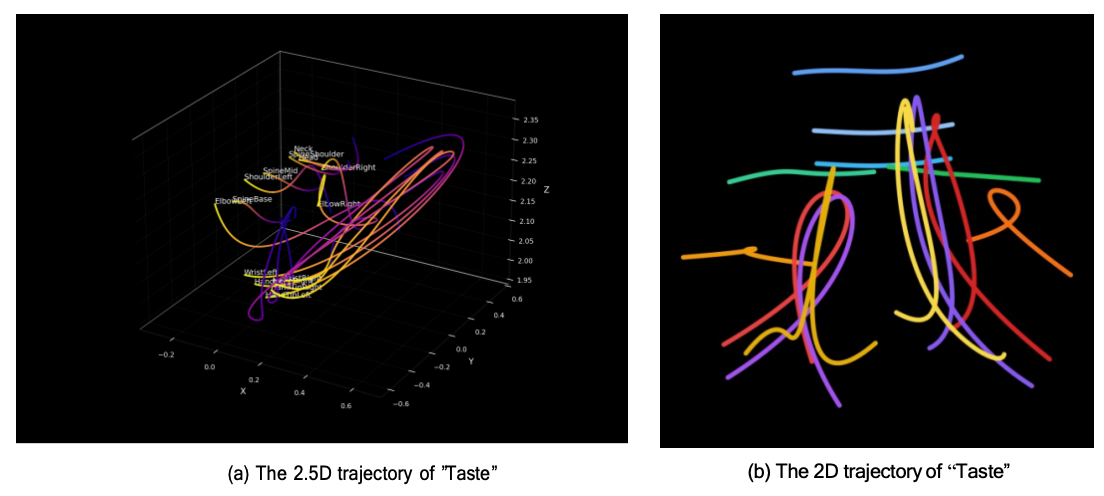}
\end{center}

Fig. 3. Representative trajectories shown as an trained movers2.5D
rendering and a 2D plan-view path (derived from the original motion
data).

Finally, we obtained the smell--verb--intent--trajectory mapping
summarized in Fig. 4 (full version in the Appendix).

\begin{center}
\includegraphics[width=\linewidth,height=0.9\textheight,keepaspectratio]{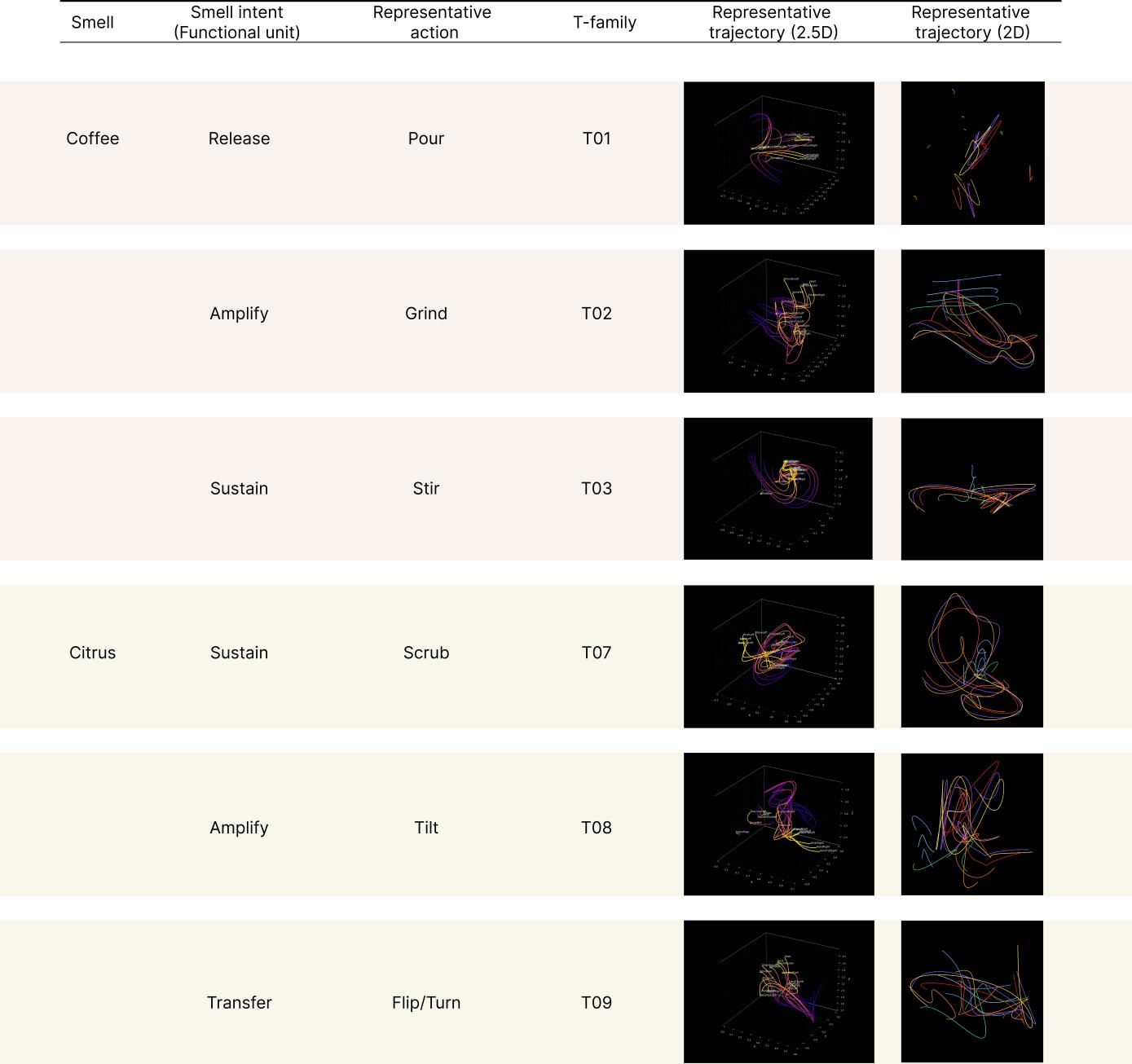}
\end{center}

Fig. 4. Smell-verb-intent-trajectory mapping (see Appendix
\hyperref[appendix]{A} for the full version).

\section{5 PHASE 3: TRAJECTORY→GESTURE}\label{phase-3-trajectorygesture}

\subsection{5.1 Aim}\label{aim-2}

Building on the Phase 1--2 smell--Verb Lexicon and smell--Action
Trajectory Set, this phase explores the transferability and ideation
potential of these intermediate assets through a structured design
workshop, generating preliminary evidence on clarity, learnability, and
generativity. We adopt scenario-based design, experience prototyping,
and soma/embodied ideation, and follow storyboarding best practices to
support team convergence and consistent expression(Truong et al. 2006;
Buchenau and Suri 2000; Carroll 2003; Hook 2018). We do not re-annotate
Phases 1/2; instead, we directly reuse two intermediate tables:

Table A: smell--Verb--Intent (Phase 1)

Table B: smell--Verb--Intent--Trajectory (Phase 2; with T-family labels)

The goal is to translate trajectories into gesture card representations
and conduct a formative evaluation of the gesture cards.

\subsection{5.2 Participants \& Setup}\label{participants-setup}

\textbf{Participants.} Designers (N=8) were UX/IxD professionals with
3--7 years of practice. None had prior smell-related project experience,
but all had previously worked on embodied-cognition--informed design
projects. Student raters (N=12) were second- or third-year master's
design students who conducted a blind formative evaluation.

\textbf{Materials.} We prepared Phase 1/2 summary sheets (one per smell)
listing smell-operational intents, representative verbs, and a
recommended trajectory, plus a trajectory reference set (T01--T13) as
path-trace visualizations to support replication and discussion. Gesture
concepts were recorded in a standardized card format linking gesture
paths to intent and movement descriptor.

.

\subsection{5.3 Workshop Procedure}\label{workshop-procedure}

\textbf{Step1 Familiarization.} Participants reviewed the reference
trajectories (T01--T13) with brief cues on direction, tempo, effort, and
key LMA descriptors, then air-traced the paths with their dominant hand
to internalize the movement.

\textbf{Step2 Reproduction and capture.} Participants learned and
reproduced the hand-only gestures from the Phase 2 table (2--3 takes
each), recorded using three synchronized fixed cameras (front, side, and
oblique/overhead).

\textbf{Step3 Trace alignment.} We extracted 2D path traces from each
view and aligned them to a shared coordinate canvas to support
consistent comparison and documentation.

\textbf{Step4 Documentation.} A representative trajectory was documented
in the gesture-card format and refined for clarity and reuse.

\textbf{Step5 Formative evaluation.} Graduate design students (N = 12)
blindly rated the gesture cards on Generativity, Clarity, Learnability,
Comfort, and Memorability using a 7-point Likert scale (Vatavu \&
Wobbrock, 2015; Wobbrock et al., 2009). We applied a Friedman test with
Wilcoxon signed-rank post hoc comparisons and Holm correction. (**p
\textless{} .01, *p \textless{} .05, n.s.).

\begin{center}
\includegraphics[width=0.49\linewidth,height=0.9\textheight,keepaspectratio]{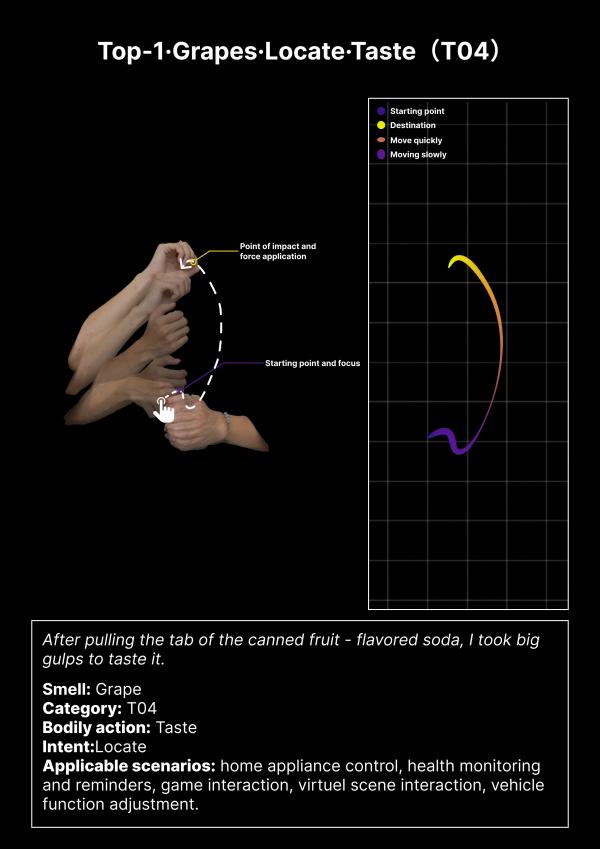}%
\hfill
\includegraphics[width=0.49\linewidth,height=0.9\textheight,keepaspectratio]{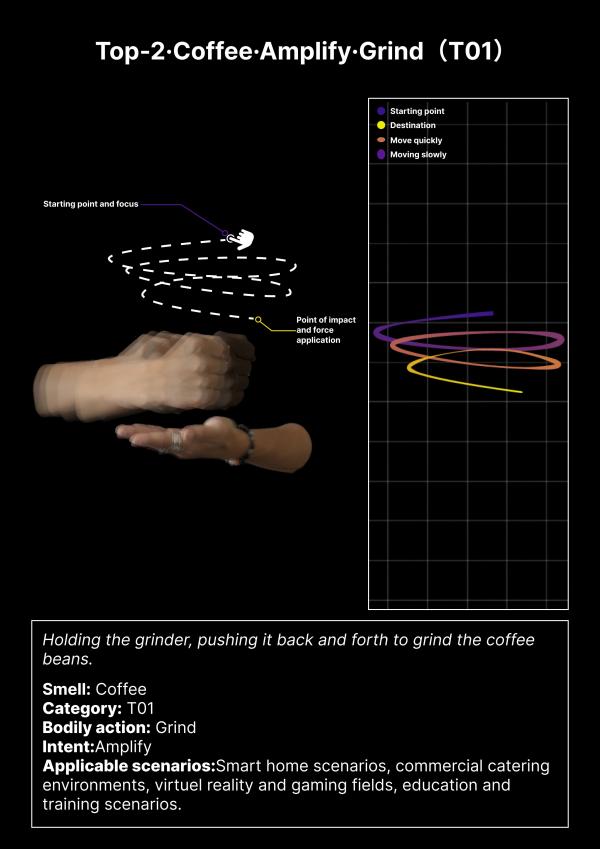}
\end{center}

(a) Top-1: Grapes•Locate•Taste (T04) (b) Top-2: Coffee•Amplify•Grind
(T01)

Fig. 5. Representative gesture cards for
smell--intent--action--trajectory mappings.

\subsection{5.4 Key Findings}\label{key-findings}

\emph{Gesture cards in use.} Designers used the Trajectory Book
(T01--T13) as a kind of ``body lexicon'' and, under minimal scenario
constraints, moved through a short sequence of actions: view a
trajectory, replicate it with their own bodies, lightly modulate or
concatenate it, and then finalise a 2D gesture sketch. Each Gesture Card
combines an animated or static 2.5D rendering with a simplified 2D path
and a brief movement description. For example, coffee---release (pour,
T02) converges to``the right hand tilts outward-- down from the chest
and returns''; grape---locate (taste, T04) reads as ``hand approaches
the mouth, makes a short rotation, then dwells''; lavender---sustain
(rotate, T03) is consistently drawn as ``a gentle circular wrist
motion.'' Each card also records Interpret / Smell / Category / Bodily
action / Intent / Applicable scenarios, mapped one-to-one to the Phase 2
Table 3 entries to support teaching, retrieval, and reuse.

\emph{Formative evaluation.} Blind ratings were positive overall
(7-point, N=12): Clarity (Mdn = 6.0, p \textless{} .01), Generativity
(Inspiration) (≈ 5.7, p \textless{} .05), and Learnability (≈ 5.5, p
\textless{} .05) were significantly above the neutral point (= 4),
whereas Comfort (≈ 5.0) and Memorability (≈ 4.4) were n.s. Comments
echoed these trends: ``The release one clearly feels like pouring it out
(coffee---release/pour, T02).'' ``\,'Peel' is easy to pick up
(grape---peel).'' ``I'm not sure I'd use it right away, but it suggests
I wouldn't need to work hard to invent lots of gestures.'' These
findings are interpreted as preliminary indicators of representational
quality, rather than as evidence of adoption or effectiveness within
sustained design practice. Longitudinal evaluation in studio and
classroom settings constitutes a clear direction for future work.

\section{6 Discussion}\label{discussion}

Over the past decade, olfactory interaction research has progressed from
prototypes and toolkits asking "can we emit smell, and how?" to a
systematic inquiry into how smell couples with interaction structure and
design practice (Bødker 2006; Bødker 2015; Maggioni et al. 2020; Obrist
et al. 2016). Hardware and co-creation efforts have moved olfaction from
a peripheral sense to a designable medium, with recent work such as
Smell Above All (Beşevli et al. 2024) exploring smell-centric futures
and situated interaction possibilities. Building on these insights, we
shift focus from device capability to reusable design representations
around body--intent--trajectory:when smell is perceived, what is the
desired action of smell (operational intent), and how can that intent be
expressed, explored, learned, and shared through movement trajectories
and interaction gestures. Smell exhibits recurring perception--action
coupling tendencies, which grounds our use of an embodied-cognition
framework for study conception and method design (Bò et al. 2024;
Iravani et al. 2021).The contribution of this work is therefore located
at the level of design representations, structured reusable artifacts
that serve as scaffolds for ideation, teaching, and communication,
rather than at the level of validated design methods whose effectiveness
in practice has been empirically established.

\textbf{Phase 1(Semantics): introduces a smell--scenario--intent--verb
framework.} We extract high-frequency verbs from real-world user
narratives, then abstract them into operational intents of smell and
align those intents with canonical scenarios (story pack). The value is
not to ``define smell,'' but to align team language: first decide what
change we want smell to effect, then select (or craft) the action---thus
avoiding pose-first, meaning-later practices and keeping the link
between action intent and smell grounded in users'stories. To ensure
semantic consistency and traceability, we combine thematic analysis with
affinity diagramming and organize design around scenarios (Braun and
Clarke 2006; Carroll 2003; Holtzblatt and Beyer 1997; Maggioni et al.
2020; Obrist et al. 2014).

\textbf{Phase 2 (Body): grounds intent in motion trajectories and, from
geometry, inductively derives smell--trajectory families.} Participants
from diverse backgrounds used a smell--verb root list to create actions
and articulate intents, producing trajectory data. We processed and
analyzed these data through geometric alignment (DTW/Procrustes) and
semantic consistency checks (dual LMA coding), converging on a
tripartite correspondence among smell, action intent, and trajectory,
summarized as correspondence tables (Bartenieff and Lewis 2013; Dryden
and Mardia 2016; Guest 2013; Müller 2007; Rousseeuw 1987; Vatavu and
Wobbrock 2015; Wobbrock et al. 2009).

\textbf{Phase 3 (Design): distills the Phase 1/2 smell--trajectory
results into reusable gesture assets.} Designers first view and
replicate trajectories, then apply minimal modulation, and finally
package gesture, intent, trajectory ID, LMA tags, and movement
descriptor into gesture cards. This pipeline converts the explicit
mappings and representations from Phases 1 and 2 into teachable and
assessable design representations (Buchenau and Suri 2000; Carroll 2003;
Hook 2018; Maggioni et al. 2019; Vatavu and Wobbrock 2015; Wobbrock et
al. 2009).The choice of gesture as the primary representational form
reflects three converging rationales. Smell resists precise verbal
encoding (Majid 2021), making purely semantic representations
insufficient. Smell-action coupling operates largely before conscious
awareness --- raw embodied responses such as recoiling or reaching are
too immediate to be directly usable in design, and must be translated
into discrete, learnable forms (Iravani et al. 2021). Gesture, as
simulated action, preserves spatial trajectory, temporal rhythm, and
effort quality simultaneously --- dimensions that semantic labels or
spatial diagrams capture only partially (Hostetter and Alibali 2008;
McNeill 1992). Within design practice, gesture elicitation constitutes
an established pathway for externalizing embodied knowledge (Wobbrock et
al. 2009; Vatavu and Wobbrock 2015), and gesture cards as
intermediate-level knowledge serve as scaffolds between experiential
insight and reusable design resource (Höök and Löwgren 2012).

\textbf{Future work.} Future work will test the stability of these
mappings across populations, cultures, and settings through participant
validation protocols and inter-rater reliability testing, and examine
how intent and trajectory choices change with task demands and
multisensory context (Brooks et al. 2023; Vatavu and Wobbrock 2016;
Wobbrock et al. 2009). We will also extend the assets to place-based and
service contexts, where smell can support narrative experiences and
orientation through situated cues (Henshaw 2013; Beşevli et al. 2024;
Bembibre and Strlič 2017; Schwarz and Hamburger 2024; Perkins and McLean
2020).

\section{7 CONCLUSION}\label{conclusion}

As an exploratory study, this work translates ``what people want to do
when they smell something'' into an operational chain from intent to
trajectory to gesture. Following a smell, intent, trajectory, gesture
pathway, we organize smell semantics into an exploratory
intent--trajectory--gesture pipeline: (1) a lexicon and an intent-first
principle that support early-stage concept generation; (2) a shared
vocabulary for cross-role collaboration (gesture cards with GIF
animations, LMA labels, and movement descriptors); and (3) structured
representations that make embodied experience more shareable and
comparable. We argue that this body-first approach complements hardware
prototyping and scenario co-creation, enabling replication, extension,
and evaluation (Feng et al. 2019).

\subsection{Limitations}\label{limitations}

\textbf{Smell coverage limitation.} We focus on four categories of
everyday smells. Each category uses one commercial sample, so stability
across intensity gradients, blends, and temporal profiles remains to be
tested.\\
\textbf{Formative evaluation and bodily computation.} Phase 3 is
formative, and Phase 2 prioritizes upper-limb trajectories following a
geometric decoupling rationale. Future work will test larger samples and
gradually incorporate trunk and lower-limb effort measures.

\textbf{Interpretive stability across translation stages.} The pipeline
involves successive re-encodings from smell experience to gesture
representation. Each stage introduces interpretive choices and potential
information loss: verb extraction reflects
participants\textquotesingle{} linguistic and cultural framings (Majid
2021; Braun and Clarke 2006); k-means clustering imposes discrete
boundaries on continuous movement (Rousseeuw 1987); and 2D rendering
simplifies 3D geometry in favour of communicability. We treat these as
inherent tensions of exploratory work rather than resolvable errors, and
acknowledge that mapping stability across users and contexts remains an
open empirical question to be addressed through participant validation
and inter-rater reliability testing.

\section{}\label{section}

\section{}\label{section-1}

\section{}\label{section-2}

\section{}\label{section-3}

\section{}\label{section-4}

\section{Appendix}\label{appendix}

\begin{center}
\includegraphics[width=\linewidth,height=0.9\textheight,keepaspectratio]{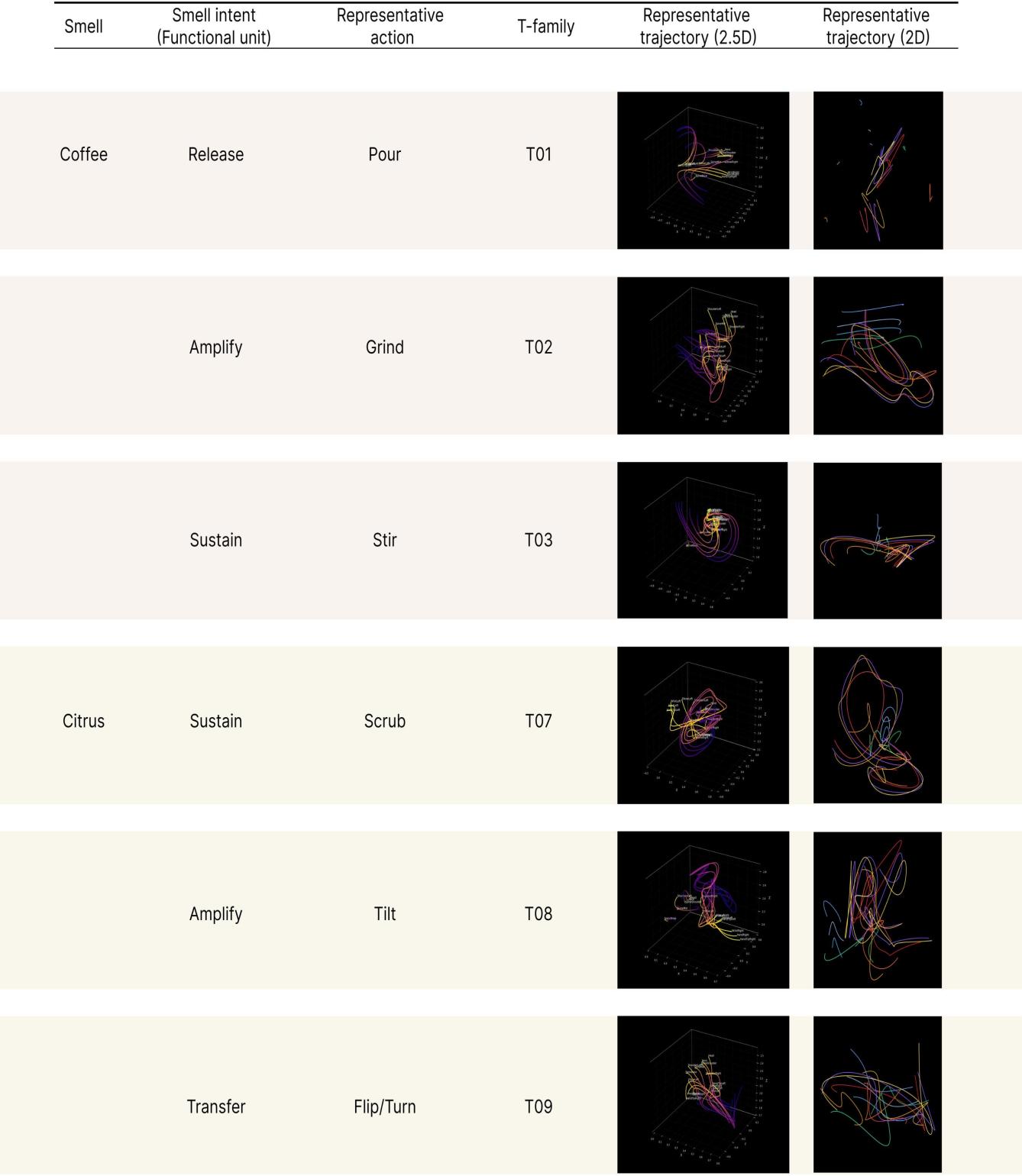}
\end{center}

\begin{center}
\includegraphics[width=\linewidth,height=0.9\textheight,keepaspectratio]{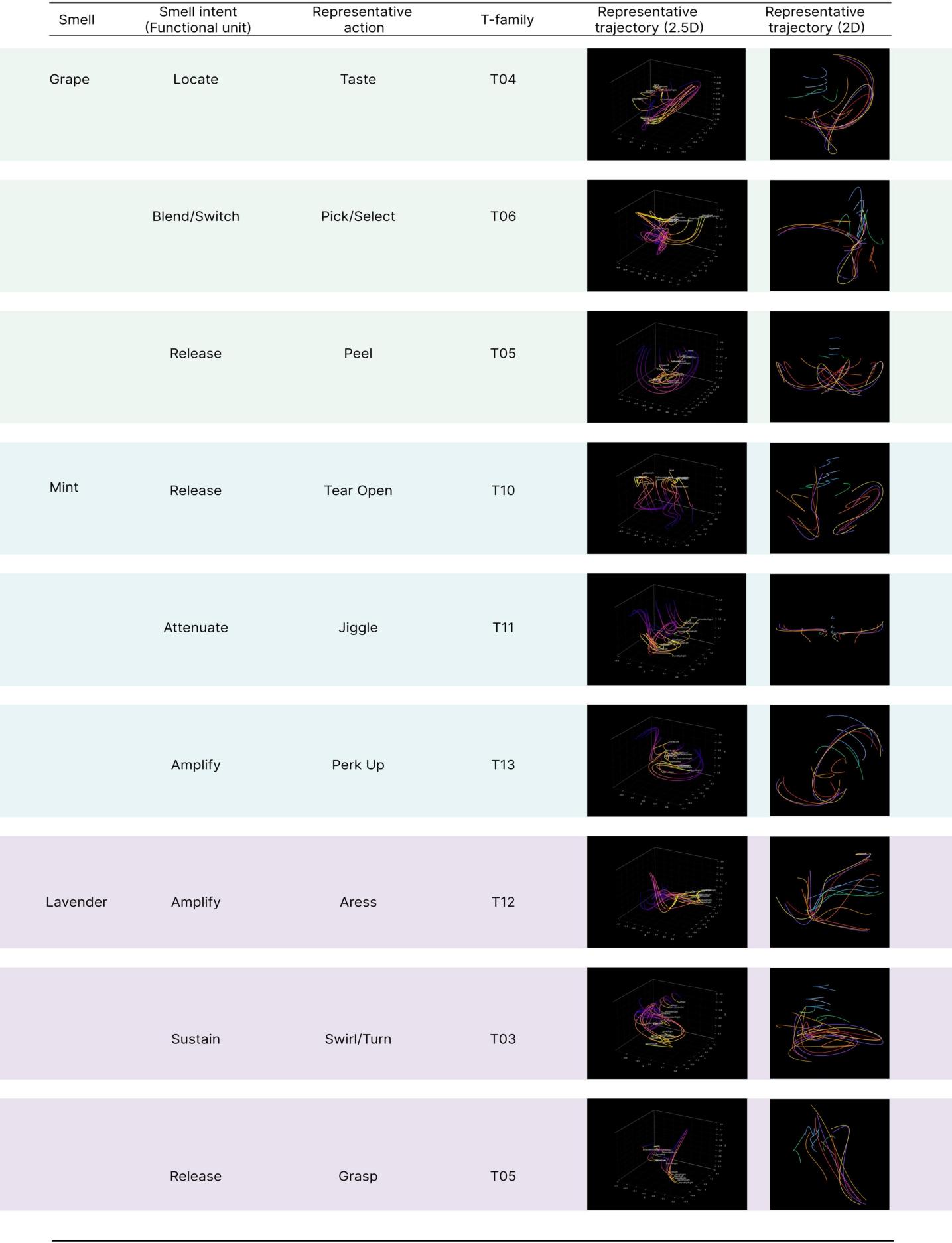}
\end{center}

Full image of Fig. 4 : smell--intent--action--trajectory mappings with
representative trajectories.

\textbf{Funding}\\
Ministry of Education of the People's Republic of China (No. 2021160042,
2021). Humanities and Social Sciences Foundation of Hubei Province (No.
HBSK244035).

\textbf{Disclosure statement}

No potential conﬂict of interest was reported by the author(s).

\section{References}\label{references}

Amores, J., and P. Maes. 2017. ``Essence: Olfactory Interfaces for
Unconscious Influence of Mood and Cognitive Performance.'' In
Proceedings of the 2017 CHI Conference on Human Factors in Computing
Systems, 28--34.

Aruin, A. S., and M. L. Latash. 1995. ``Directional Specificity of
Postural Muscles in Feed-Forward Postural Reactions during Fast
Voluntary Arm Movements.'' Experimental Brain Research 103 (2):
323--332.

Barsalou, L. W. 2008. ``Grounded Cognition.'' Annual Review of
Psychology 59 (1): 617--645.

Bartenieff, I., and D. Lewis. 2013. Body Movement: Coping with the
Environment. Routledge.

Bembibre, C., and M. Strlič. 2017. ``Smell of Heritage: A Framework for
the Identification, Analysis and Archival of Historic Odours.'' Heritage
Science 5: 1.

Beşevli, C., G. Brianza, C. Dawes, N. Shabanova, S. Mathur, M. Lechner,
E. Maggioni, D. Boak, C. Philpott, A. F. G. Schieck, et al. 2024.
``Smell Above All: Envisioning Smell-Centred Future Worlds.'' In
Proceedings of the 2024 ACM Designing Interactive Systems Conference,
2530--2544.

Bødker, S. 2006. ``When Second Wave HCI Meets Third Wave Challenges.''
In Proceedings of the 4th Nordic Conference on Human-Computer
Interaction: Changing Roles, 1--8.

Bødker, S. 2015. ``Third-Wave HCI, 10 Years Later---Participation and
Sharing.'' Interactions 22 (5): 24--31.

Bò, E. D., C. Cecchetto, A. L. Callara, A. Greco, F. Mura, N. Vanello,
F. Di Francesco, E. P. Scilingo, and C. Gentili. 2024. ``Emotion
Perception through the Nose: How Olfactory Emotional Cues Modulate the
Perception of Neutral Facial Expressions in Affective Disorders.''
Translational Psychiatry 14 (1): 342.

Braun, V., and V. Clarke. 2006. ``Using Thematic Analysis in
Psychology.'' Qualitative Research in Psychology 3 (2): 77--101.

Brooks, J., and P. Lopes. 2023. ``Smell and Paste: Low-Fidelity
Prototyping for Olfactory Experiences.'' In Proceedings of the 2023 CHI
Conference on Human Factors in Computing Systems, 1--16.

Brooks, J., P. Lopes, M. Obrist, J. A. Fernandez, and J. Kaye. 2023.
``Third Wave or Winter? The Past and Future of Smell in HCI.'' In
Extended Abstracts of the 2023 CHI Conference on Human Factors in
Computing Systems, 1--4.

Buchenau, M., and J. F. Suri. 2000. ``Experience Prototyping.'' In
Proceedings of the 3rd Conference on Designing Interactive Systems:
Processes, Practices, Methods, and Techniques, 424--433.

Cao, Z., T. Simon, S.-E. Wei, and Y. Sheikh. 2017. ``Realtime
Multi-Person 2D Pose Estimation Using Part Affinity Fields.'' In
Proceedings of the IEEE Conference on Computer Vision and Pattern
Recognition, 7291--7299.

Carroll, J. M. 2003. Making Use: Scenario-Based Design of Human-Computer
Interactions. MIT Press.

Clark, R. A., B. F. Mentiplay, E. Hough, and Y. H. Pua. 2019.
``Three-Dimensional Cameras and Skeleton Pose Tracking for Physical
Function Assessment: A Review of Uses, Validity, Current Developments
and Kinect Alternatives.'' Gait \& Posture 68: 193--200.

Clark, R. A., Y.-H. Pua, K. Fortin, C. Ritchie, K. E. Webster, L.
Denehy, and A. L. Bryant. 2012. ``Validity of the Microsoft Kinect for
Assessment of Postural Control.'' Gait \& Posture 36 (3): 372--377.

Cohen, J. 1960. ``A Coefficient of Agreement for Nominal Scales.''
Educational and Psychological Measurement 20 (1): 37--46.

Coppola, D. M., and B. Slotnick. 2018. ``Odor-Cued Bitter Taste
Avoidance.'' Chemical Senses 43 (4): 239--247.

Dematte, M. L., D. Sanabria, R. Sugarman, and C. Spence. 2006.
``Cross-Modal Interactions between Olfaction and Touch.'' Chemical
Senses 31 (4): 291--300.

Dmitrenko, D., E. Maggioni, C. T. Vi, and M. Obrist. 2017. ``What Did I
Sniff? Mapping Scents onto Driving-Related Messages.'' In Proceedings of
the 9th International Conference on Automotive User Interfaces and
Interactive Vehicular Applications, 154--163.

Dobbelstein, D., S. Herrdum, and E. Rukzio. 2017. ``inScent: A Wearable
Olfactory Display as an Amplification for Mobile Notifications.'' In
Proceedings of the 2017 ACM International Symposium on Wearable
Computers, 130--137.

Dourish, P. 2001. Where the Action Is: The Foundations of Embodied
Interaction. MIT Press.

Dryden, I. L., and K. V. Mardia. 2016. Statistical Shape Analysis: With
Applications in R. John Wiley \& Sons.

Feng, A., Y. V. Chen, R. Li, and L. Ding. 2019. ``How Smell Can Help
Visually Impaired in Health and Well-Being--A Cognitive Experiment.''
The Design Journal 22: 371--386.

Flavián, C., S. Ibáñez-Sánchez, and C. Orús. 2021. ``The Influence of
Scent on Virtual Reality Experiences: The Role of Aroma-Content
Congruence.'' Journal of Business Research 123: 289--301.

Greenwald, A. G., D. E. McGhee, and J. L. Schwartz. 1998. ``Measuring
Individual Differences in Implicit Cognition: The Implicit Association
Test.'' Journal of Personality and Social Psychology 74 (6): 1464.

Greifeneder, R., H. Bless, and M. T. Pham. 2011. ``When Do People Rely
on Affective and Cognitive Feelings in Judgment? A Review.'' Personality
and Social Psychology Review 15 (2): 107--141.

Guess, T. M., R. Bliss, J. B. Hall, and A. M. Kiselica. 2022.
``Comparison of Azure Kinect Overground Gait Spatiotemporal Parameters
to Marker Based Optical Motion Capture.'' Gait \& Posture 96: 130--136.

Guest, A. H. 2013. Labanotation: The System of Analyzing and Recording
Movement. Routledge.

Hanson-Vaux, G., A.-S. Crisinel, and C. Spence. 2013. ``Smelling Shapes:
Crossmodal Correspondences between Odors and Shapes.'' Chemical Senses
38 (2): 161--166.

Henshaw, V. 2013. Urban Smellscapes: Understanding and Designing City
Smell Environments. Routledge.

Herz, R. S. 2016. ``The Role of Odor-Evoked Memory in Psychological and
Physiological Health.'' Brain Sciences 6 (3): 22.

Holtzblatt, K., and H. Beyer. 1997. Contextual Design: Defining
Customer-Centered Systems. Elsevier.

Hook, K. 2018. Designing with the Body: Somaesthetic Interaction Design.
MIT Press.

Höök, K., and J. Löwgren. 2012. ``Strong Concepts: Intermediate-Level
Knowledge in Interaction Design Research.'' ACM Transactions on
Computer-Human Interaction (TOCHI) 19 (3): 1--18.

Hummel, T., B. Sekinger, S. R. Wolf, E. Pauli, and G. Kobal. 1997.
``\,`Sniffin' Sticks': Olfactory Performance Assessed by the Combined
Testing of Odor Identification, Odor Discrimination and Olfactory
Threshold.'' Chemical Senses 22 (1): 39--52.

Iravani, B., M. Schaefer, D. A. Wilson, A. Arshamian, and J. N.
Lundström. 2021. ``The Human Olfactory Bulb Processes Odor Valence
Representation and Cues Motor Avoidance Behavior.'' Proceedings of the
National Academy of Sciences 118 (42).

Jacobs, L. F. 2023. ``The PROUST Hypothesis: The Embodiment of Olfactory
Cognition.'' Animal Cognition 26 (1): 59--72.

Kaye, J. J. 2004. ``Making Scents: Aromatic Output for HCI.''
Interactions 11 (1): 48--61.

Lei, Y., Q. Lu, and Y. Xu. 2022. ``O\&O: A DIY Toolkit for Designing and
Rapid Prototyping Olfactory Interfaces.'' In Proceedings of the 2022 CHI
Conference on Human Factors in Computing Systems, 1--21.

Li, J., Y. Wang, Z. Cui, J. Brooks, Y. Yan, Z. Lou, and Y. Li. 2025.
``Mid-Air Gestures for Proactive Olfactory Interactions in Virtual
Reality.'' In Proceedings of the 2025 CHI Conference on Human Factors in
Computing Systems, 1--18.

Li, W., J. D. Howard, T. B. Parrish, and J. A. Gottfried. 2008.
``Aversive Learning Enhances Perceptual and Cortical Discrimination of
Indiscriminable Odor Cues.'' Science 319 (5871): 1842--1845.

Lin, Y. C., S.-L. Yu, A.-Y. Zhuang, C. Lee, Y. A. Ting, S.-K. Lee, B.-J.
Lin, and P.-C. Kuo. 2024. ``Representing Scents: An Evaluation Framework
of Scent-Related Experiences through Associations between Grounded and
Psychophysiological Data.'' International Journal of Human-Computer
Studies 192: 103357.

Maggioni, E., R. Cobden, D. Dmitrenko, K. Hornbæk, and M. Obrist. 2020.
``SMELL SPACE: Mapping Out the Olfactory Design Space for Novel
Interactions.'' ACM Transactions on Computer-Human Interaction (TOCHI)
27 (5): 1--26.

Maggioni, E., R. Cobden, and M. Obrist. 2019. ``OWidgets: A Toolkit to
Enable Smell-Based Experience Design.'' International Journal of
Human-Computer Studies 130: 248--260.

Majid, A. 2021. ``Human Olfaction at the Intersection of Language,
Culture, and Biology.'' Trends in Cognitive Sciences 25 (2): 111--123.

McHugh, M. L. 2012. ``Interrater Reliability: The Kappa Statistic.''
Biochemia Medica 22 (3): 276--282.

McLean, K. 2017. ``Smellmap: Amsterdam---Olfactory Art and Smell
Visualization.'' Leonardo 50 (1): 92--93.

Mehrotra, S., A. Brocker, M. Obrist, and J. Borchers. 2022. ``The Scent
of Collaboration: Exploring the Effect of Smell on Social
Interactions.'' In CHI Conference on Human Factors in Computing Systems
Extended Abstracts, 1--7.

Müller, M. 2007. Information Retrieval for Music and Motion. Springer.

Niedenthal, S., W. Fredborg, P. Lundén, M. Ehrndal, and J. K. Olofsson.
2023. ``A Graspable Olfactory Display for Virtual Reality.''
International Journal of Human-Computer Studies 169: 102928.

Obrist, M., A. N. Tuch, and K. Hornbaek. 2014. ``Opportunities for Odor:
Experiences with Smell and Implications for Technology.'' In Proceedings
of the SIGCHI Conference on Human Factors in Computing Systems,
2843--2852.

Obrist, M., C. Velasco, C. Vi, N. Ranasinghe, A. Israr, A. Cheok, C.
Spence, and P. Gopalakrishnakone. 2016. ``Sensing the Future of HCI:
Touch, Taste, and Smell User Interfaces.'' Interactions 23 (5): 40--49.

Oleszkiewicz, A., K. Pettke, H. Olze, F. C. Uecker, and T. Hummel. 2023.
``Effects of Odors on Posture.'' Journal of Sensory Studies 38 (1):
e12796.

Perkins, C., and K. McLean. 2020. ``Smell Walking and Mapping.'' In
Mundane Methods, 156--173. Manchester University Press.

Poo, C., G. Agarwal, N. Bonacchi, and Z. F. Mainen. 2022. ``Spatial Maps
in Piriform Cortex during Olfactory Navigation.'' Nature 601 (7894):
595--599.

Puska, P., S. Kurki, M. Lähdesmäki, M. Siltaoja, and H. Luomala. 2018.
``Sweet Taste of Prosocial Status Signaling: When Eating Organic Foods
Makes You Happy and Hopeful.'' Appetite 121: 348--359.

Rousseeuw, P. J. 1987. ``Silhouettes: A Graphical Aid to the
Interpretation and Validation of Cluster Analysis.'' Journal of
Computational and Applied Mathematics 20: 53--65.

Schreuder, E., M. R. Hoeksma, M. A. Smeets, and G. R. Semin. 2014. ``The
Effects of Odor and Body Posture on Perceived Duration.'' Frontiers in
Neurorobotics 8: 6.

Schwarz, M., and K. Hamburger. 2024. ``Memory Effects of Visual and
Olfactory Landmark Information in Human Wayfinding.'' Cognitive
Processing 25 (1): 37--51.

Schwarz, N., and G. L. Clore. 1983. ``Mood, Misattribution, and
Judgments of Well-Being: Informative and Directive Functions of
Affective States.'' Journal of Personality and Social Psychology 45 (3):
513.

Sebanz, N., and M. Shiffrar. 2009. ``Detecting Deception in a Bluffing
Body: The Role of Expertise.'' Psychonomic Bulletin \& Review 16 (1):
170--175.

Shotton, J., A. Fitzgibbon, M. Cook, T. Sharp, M. Finocchio, R. Moore,
A. Kipman, and A. Blake. 2011. ``Real-Time Human Pose Recognition in
Parts from Single Depth Images.'' In CVPR 2011, 1297--1304.

Slovic, P., M. L. Finucane, E. Peters, and D. G. MacGregor. 2007. ``The
Affect Heuristic.'' European Journal of Operational Research 177 (3):
1333--1352.

Spence, C. 2020. ``Olfactory-Colour Crossmodal Correspondences in Art,
Science, and Design.'' Cognitive Research: Principles and Implications 5
(1): 52.

Springer, S., and G. Y. Seligmann. 2016. ``Validity of the Kinect for
Gait Assessment: A Focused Review.'' Sensors 16: 2.

Truong, K. N., G. R. Hayes, and G. D. Abowd. 2006. ``Storyboarding: An
Empirical Determination of Best Practices and Effective Guidelines.'' In
Proceedings of the 6th Conference on Designing Interactive Systems,
12--21.

Vatavu, R.-D., and J. O. Wobbrock. 2015. ``Formalizing Agreement
Analysis for Elicitation Studies: New Measures, Significance Test, and
Toolkit.'' In Proceedings of the 33rd Annual ACM Conference on Human
Factors in Computing Systems, 1325--1334.

Vatavu, R.-D., and J. O. Wobbrock. 2016. ``Between-Subjects Elicitation
Studies: Formalization and Tool Support.'' In Proceedings of the 2016
CHI Conference on Human Factors in Computing Systems, 3390--3402.

Vatavu, R.-D., and J. O. Wobbrock. 2022. ``Clarifying Agreement
Calculations and Analysis for End-User Elicitation Studies.'' ACM
Transactions on Computer-Human Interaction (TOCHI) 29 (1): 1--70.

von Hippel, E. 1986. ``Lead Users: A Source of Novel Product Concepts.''
Management Science 32 (7): 791--805.

Wang, Y., J. Amores, and P. Maes. 2020. ``On-Face Olfactory
Interfaces.'' Association for Computing Machinery, New York, NY, 1--9.

Wang, Y., Z. Cui, H. Gong, and T. Chen. 2024. ``Olfackit: A Toolkit for
Integrating Atomization-Based Olfactory Interfaces into Daily
Scenarios.'' International Journal of Human--Computer Interaction 40
(16): 4392--4411.

Ward, R. J., S. M. Wuerger, and A. Marshall. 2022. ``Smelling
Sensations: Olfactory Crossmodal Correspondences.'' Journal of
Perceptual Imaging 5 (0): 000402.

Wilson, M. 2002. ``Six Views of Embodied Cognition.'' Psychonomic
Bulletin \& Review 9 (4): 625--636.

Wobbrock, J. O., and J. A. Kientz. 2016. ``Research Contributions in
Human-Computer Interaction.'' Interactions 23 (3): 38--44.

Wobbrock, J. O., M. R. Morris, and A. D. Wilson. 2009. ``User-Defined
Gestures for Surface Computing.'' In Proceedings of the SIGCHI
Conference on Human Factors in Computing Systems, 1083--1092.

Hostetter, A. B., and M. W. Alibali. 2008. "Visible Embodiment: Gestures
as Simulated Action." Psychonomic Bulletin and Review 15 (3): 495--514.

McNeill, D. 1992. Hand and Mind: What Gestures Reveal about Thought.
Chicago: University of Chicago Press.

Zimmerman, J., J. Forlizzi, and S. Evenson. 2007. "Research Through
Design as a Method for Interaction Design Research in HCI." In
Proceedings of the SIGCHI Conference on Human Factors in Computing
Systems, 493--502.

\end{document}